\documentclass[aps,pra,reprint,groupedaddress,floatfix]{revtex4-2}

\usepackage[utf8]{inputenc}
\usepackage[english]{babel}
\usepackage[T1]{fontenc}
\usepackage{amsmath,amssymb}
\usepackage{physics}
\usepackage{mathtools,mathrsfs}
\usepackage{amsbsy}
\usepackage{hyperref}
\usepackage{tikz}
\usetikzlibrary{arrows.meta,decorations.pathreplacing,positioning}
\usetikzlibrary{shapes.geometric} 
\usepackage{braket}
\usepackage{tikz-cd}
\usepackage{qcircuit}
\usepackage[caption = false]{subfig}
\usepackage{xcolor}
\definecolor{revisionblue}{RGB}{0,87,184}
\definecolor{REVISIONBLUE}{RGB}{0,87,184}

\definecolor{revisionred}{RGB}{190,8,1}

\usepackage{orcidlink}

\graphicspath{{./Figures/}}

\begin{document}

\title{Sublattice Entanglement of a Ladder with Mutual Statistical Phases}

\author{Babatunde M. Ayeni \orcidlink{0000-0003-4035-149X} }
\email{babatunde.ayeni@tud.ie}
\affiliation{School of Mathematics and Statistics, Technological University Dublin, Ireland}
\affiliation{Department of Physics, National University of Ireland, Maynooth, Ireland}

\date{\today}

\begin{abstract}
When a nonlocal unitary transformation removes a statistical phase from a lattice Hamiltonian, the energy spectrum remains unchanged, yet the phase persists in the eigenstates and can generate entanglement between subsystems. Here, we study how such a phase entangles the two legs of a hard-core boson ladder, the bosonic form of an integrable spin ladder with anyon-like mutual statistics, in which hopping along each leg acquires a density-dependent Peierls phase set by the other leg. A diagonal unitary decouples the legs: for open boundaries the spectrum and occupation statistics do not depend on the statistical angle, and for periodic boundaries the transformation only adds boundary fluxes. The inter-leg sublattice entanglement is therefore generated by the unitary alone. Tracing out one leg yields an exact random-unitary channel on the remaining leg, whose matrix elements are the multivariate full counting statistics (FCS) of the traced leg with counting fields fixed by the statistical phase. Being random-unitary, the channel is completely positive and limits the generated entropy to the Shannon entropy of the traced leg's occupations. We also show that the cumulant expansion of the FCS separates a unitary rotation from a completely positive correlated dephasing, and that truncations beyond second order need not be positive. For a pure product input, the purity follows exactly from the occupation statistics of the two legs, with leading loss set by the overlap of their fluctuations. Exact diagonalization up to eight rungs confirms the decoupling, channel, cumulant expansion, and purity to high precision. The reduced density matrix is thus given exactly by counting statistics, without assuming a Gaussian reduced state or a quadratic entanglement Hamiltonian.
\end{abstract}

\maketitle

\section{Introduction}
\label{sec:intro}
In two dimensions, exchange statistics is not limited to bosons and fermions: exchanging two anyons multiplies the wave function by a phase $e^{i\theta}$ with arbitrary $\theta$ \cite{LeinaasMyrheim1977,GoldinMenikoffSharp1981,Wilczek1982}. Different species of particles of an anyonic model can also carry mutual statistics, for example the electric and magnetic charges of the toric code. Their effective theory is a mutual Chern--Simons theory \cite{Kitaev2003,KouLevinWen2008}. On a lattice, statistical phases are implemented by Jordan--Wigner strings, phase operators that depend on the positions of the other particles \cite{JordanWigner1928,LiebSchultzMattis1961,Fradkin1989,ChenKapustinRadicevic2018}; in one dimension, such a string counts the particles on one side of a site. A fractional Jordan--Wigner transformation maps lattice anyons onto bosons whose hopping carries a phase that depends on the local occupation \cite{Keilmann2011,GreschnerSantos2015,Tang2015,ArcilaForero2016,Bonkhoff2021,Bonkhoff2025}. Such density-dependent Peierls phases \cite{Peierls1933} make the gauge field depend on the matter configuration \cite{Edmonds2013,Greschner2014,Bermudez2015,Chisholm2022,Aidelsburger2022} and have been realized with ultracold and Rydberg atoms \cite{Clark2018,Gorg2019,Schweizer2019,Lienhard2020,Frolian2022}.

For hard-core anyons with nearest-neighbor hopping, the string drops out of the Hamiltonian, which becomes that of hard-core bosons \cite{Hao2009,Wright2014,Tang2015}. On an open chain, the spectrum and the density then do not depend on $\theta$. The statistical angle appears instead in quantities that contain the string, such as the one-body density matrix, the momentum distribution, and the entanglement between particles \cite{Santachiara2007,SantachiaraCalabrese2008,Hao2008,Hao2009,HaoChen2012,Wright2014,Tang2015,Mani2020}.

Two-leg ladders are the simplest quasi-one-dimensional lattices on which particles can move around each other. For anyons, a ladder lets particles braid as they hop: ladders of non-Abelian anyons \cite{Nayak2008} have ground-state phases set by fusion interactions and, for mobile anyons, by braiding \cite{Poilblanc2011,Soni2016,ayeni2016simulation,Ayeni2017thesis,Ayeni2018,Kirchner2025}, and they have been studied with anyonic tensor networks \cite{singh2014matrix,Pfeifer2012,ayeni2016simulation,Kirchner2023}, beyond exact diagonalization.

Two species on a ladder can also be coupled only through statistical phases, with the hopping phase on one leg set by the occupations of the other. Such statistical coupling changes the ground-state phases of pseudofermion chains with next-nearest-neighbor hopping \cite{Agarwala2019,AdityaSen2021} and the interchain entanglement of coupled Su--Schrieffer--Heeger chains \cite{Santra2021}, and it appears in two-component anyon-Hubbard models \cite{Cardarelli2016,Chen2025}. Pozsgay et al.\ introduced an integrable zig-zag spin ladder of this type, in which particles on the same leg scatter as free fermions and particles on different legs acquire a constant phase shift \cite{Pozsgay2022}. They showed that under open boundaries a diagonal similarity transformation decouples the two legs, so that the spectrum is that of two independent XX chains. They also obtained a correlation function between the legs exactly and computed the sublattice entanglement entropy numerically, in the ground state and after a quench \cite{Pozsgay2022}.

Because the statistical phase does not manifest in the open-boundary spectrum, its effect has to be determined from the eigenstates, and entanglement provides a direct way to do so. Entanglement entropies and spectra characterize many-body states beyond local order parameters \cite{HolzheyLarsenWilczek1994,Vidal2003,CalabreseCardy2004,Amico2008,EisertCramerPlenio2010,Laflorencie2016,LiHaldane2008,CalabreseLefevre2008,Pollmann2010,Thomale2010,Chandran2014}. Across a cut between two coupled chains or layers, or across the boundary of a tensor-network state, the entanglement spectrum resembles the spectrum of an effective Hamiltonian of one part \cite{Poilblanc2010,FurukawaKim2011,PeschelChung2011,LauchliSchliemann2012,QiKatsuraLudwig2012,ChenFradkin2013,Lundgren2013,Cirac2011}. In these cases, the two parts are coupled by hopping or exchange. But it is also possible for two entangled subsystems to be coupled only through a statistical string, such as the case we consider in this paper. Entanglement across a cut does not change under unitaries that factorize across it \cite{ChenGuWen2010}, but nonlocal unitaries can change it. 

Counting statistics gives a second route to the reduced state. Full counting statistics (FCS), the probability distribution of a counted charge, was introduced for mesoscopic conductors \cite{LevitovLesovik1993,Levitov1996,Klich2003,NazarovKindermann2003,Schonhammer2007,Esposito2009}, where the counting field is the phase acquired by a detector coupled to the counted charge \cite{Levitov1996,NazarovKindermann2003}. For noninteracting fermions, the cumulants of the particle number in a region determine its entanglement entropy and spectrum \cite{KlichLevitov2009,Song2011,Song2012,CalabreseMintchevVicari2012,SusstrunkIvanov2012}, and more generally the Shannon entropy of suitable symmetry observables bounds the entanglement entropy from below \cite{KlichRefaelSilva2006}. These exact relations hold because the reduced state of free fermions is Gaussian, with a quadratic entanglement Hamiltonian \cite{ChungPeschel2001,Peschel2003,CheongHenley2004,PeschelEisler2009,CasiniHuerta2009,EislerPeschel2017,Dalmonte2022}.

These results leave open how a statistical string that can be removed from the Hamiltonian determines the reduced state across a cut that it crosses. The question matters for the ladder of Ref.~\cite{Pozsgay2022} because there the statistical angle enters neither the open-boundary spectrum nor any occupation statistics; it appears only in coherences in the occupation basis, such as those of the leg-cut reduced state. In that work, the sublattice entanglement was computed numerically, but no analytic form was given. The free-fermion relations between counting statistics and entanglement do not apply directly, because the leg-cut reduced state is generally not Gaussian even when each leg is a Slater determinant state. 

In this work we derive the exact leg-cut reduced state of this ladder, which can then be used to bound the entanglement that the string can generate. We write the model as two hard-core boson chains coupled by mutual density-dependent Peierls phases (Fig.~\ref{fig:model}). Under open boundary conditions (OBC) and up to on-site phases, it is the spin ladder of Ref.~\cite{Pozsgay2022}, with $\theta$ corresponding to $2\gamma$ in their notation. As there, a diagonal unitary $U_{\mathrm{stat}}$ maps the open ladder onto two decoupled hard-core chains, each equivalent to free fermions \cite{Girardeau1960,LiebSchultzMattis1961}, so the OBC spectrum does not depend on $\theta$. Under periodic boundary conditions (PBC), the mapping only leaves boundary fluxes on the legs, determined the statistical angle and the total number of particles on opposite legs. With either boundary condition, the leg-cut entanglement is therefore generated by $U_{\mathrm{stat}}$ alone, and it does not factorize, whereas for cuts between rungs, $U_{\mathrm{stat}}$ factorizes and leaves the Schmidt values unchanged. 

Our analysis starts from the fact that $U_{\mathrm{stat}}$ is diagonal in the occupation basis of the traced leg. It therefore acts on the retained leg as a set of unitaries $V_\mathbf{n}$ controlled by the occupation configuration $\mathbf{n}$ of the traced leg. Tracing out the leg replaces these controlled unitaries by a classical average over $\mathbf{n}$. This average is a lattice analog of a Feynman--Vernon influence functional \cite{FeynmanVernon1963,CaldeiraLeggett1983,Leggett1987,Lerose2021}, and it leads to the following results.
\begin{enumerate}
\item For every product density operator $\sigma_A\otimes\sigma_B$ of the decoupled chains, the leg-$A$ reduced state of $U_{\mathrm{stat}}(\sigma_A\otimes\sigma_B)U_{\mathrm{stat}}^\dagger$ is the random-unitary channel $\rho_A(\theta)=\sum_\mathbf{n} p_\mathbf{n} V_\mathbf{n}(\theta)\sigma_AV_\mathbf{n}^\dagger(\theta)$ [Eq.~\eqref{eq:rho_exact_fcs}], whose weights $p_\mathbf{n}$ are the occupation probabilities of the traced leg \cite{AudenaertScheel2008,Watrous2018}. The entropy generated by the string is therefore at most the Shannon entropy of $p_\mathbf{n}$ [Eq.~\eqref{eq:entropy_bounds}].
\item In the occupation basis of the retained leg, the channel multiplies $\sigma_A$ elementwise by the multivariate FCS characteristic function of the traced-leg occupations, with counting fields fixed by the string [Eq.~\eqref{eq:fcs_factor}]. For a Slater-determinant traced leg this function is an $L\times L$ determinant [Eq.~\eqref{eq:det_form}] \cite{Klich2003,Schonhammer2007}, although $\rho_A$ is generally not Gaussian.
\item In a neighborhood of zero counting field where the logarithm of the characteristic function is analytic, the channel is the exponential of commuting superoperators generated by the connected occupation cumulants [Eq.~\eqref{eq:superoperator_def}] \cite{Kubo1962}. The first cumulant gives a deterministic phase rotation, and the second gives correlated dephasing, an average over Gaussian classical phase noise [Eq.~\eqref{eq:gaussian_noise_channel}] and therefore completely positive \cite{PalmaSuominenEkert1996,BreuerPetruccione2002,Kropf2016,Chenu2017}. Although, truncations that keep third or higher cumulants need not be positive \cite{Marcinkiewicz1939,RajagopalSudarshan1974}.
\item For a pure product input, the leading purity loss and the second R\'enyi entropy are $2\theta^2\operatorname{Tr}(\kappa_B\Gamma_A)+O(\theta^4)$ [Eq.~\eqref{eq:purity_coefficient}], where $\kappa_B$ is the occupation covariance of the traced leg and $\Gamma_A$ is the covariance of the numbers of retained-leg particles to the right of each rung. Entanglement appears where these fluctuations overlap.
\end{enumerate}
Exact diagonalization for $L$ up to $8$ rungs is performed to confirm the mapping, the partial trace, the FCS and determinant formulas, and the purity coefficient (Sec.~\ref{sec:numerics}). The results give the analytic structure behind the sublattice entanglement computed numerically in Ref.~\cite{Pozsgay2022} and differ from the free-fermion relations between counting statistics and entanglement \cite{KlichLevitov2009,Song2011,Song2012} in two ways: the counting fields are fixed by the statistical string rather than introduced as free parameters, and the FCS determines the full reduced density matrix rather than only its entropy, without assuming that the reduced state is Gaussian or that its entanglement Hamiltonian is quadratic.

The quantities that enter these results can be measured with site-resolved detection. In quantum-gas microscopes, which have been used to realize one-dimensional anyons \cite{Kwan2024}, snapshots of the traced leg sample $p_n$, and the leg-cut purity $e^{-S_2}$ is measured by two-copy interference \cite{Daley2012,Islam2015,Kaufman2016} and by randomized measurements \cite{Elben2018,Brydges2019,Elben2023}. The exact reduced state also gives reference data for reduced-state tomography \cite{cramer2010efficient,Torlai2018,Huang2020,araujo2022local}, entanglement-spectrum and entanglement-Hamiltonian reconstruction \cite{Pichler2016,DalmonteVermerschZoller2018,johri2017entanglement,choo2018measurement,subacsi2019entanglement,kokail2021entanglement,mao2025sampling}, and number-conserving variational circuits \cite{gard2020efficient,anselmetti2021local,ayeni2025}.

Section~\ref{sec:model} defines the model and the statistical dressing, and Sec.~\ref{sec:fcs} derives the leg-cut influence functional and channel. Section~\ref{sec:cumulants} develops the cumulant hierarchy, Sec.~\ref{sec:modular} its consequences for the channel and the purity, and Sec.~\ref{sec:numerics} the numerical checks. Sections~\ref{sec:discussion} and \ref{sec:conclusion} give the discussion and conclusion, and the appendices the supporting derivations.

\section{Model and Gauge Transformation}
\label{sec:model}

We consider a two-leg ladder with rungs $j=1,\dots,L$ and legs $\ell\in\{A,B\}$. On each site $(j,\ell)$ we define hardcore boson operators $b_{j,\ell}$, $b_{j,\ell}^{\dagger}$, with their usual definitions, from which number operators are defined as $n_{j,\ell}=b_{j,\ell}^{\dagger} b_{j,\ell}\in\{0,1\}$. The particle numbers
\begin{equation}
N_A = \sum_{j=1}^{L} n_{j,A},
\qquad
N_B = \sum_{j=1}^{L} n_{j,B},
\end{equation}
are separately conserved. There is only intraleg hopping, but the two legs are coupled through mutual density-dependent Peierls phases adapted to a zig-zag ordering of the ladder. A convenient form of the Hamiltonian is
\begin{equation}
\label{eq:Hladder}
\begin{aligned}
H(\theta)=-t\sum_j \Bigl[&
 b_{j+1,A}^{\dagger} e^{+i\theta n_{j,B}} b_{j,A} \\
 &+b_{j+1,B}^{\dagger} e^{-i\theta n_{j+1,A}} b_{j,B}
 + \text{h.c.}\Bigr],
\end{aligned}
\end{equation}
with the obvious modification of the sum for open and periodic boundaries. The phases follow the zig-zag ordering $(1,A)<(1,B)<(2,A)<\cdots<(L,B)$: each hop acquires a phase set by the occupation of the opposite-leg site it passes in this ordering, as depicted in Fig.~\ref{fig:model}. 

The crucial microscopic fact is that, in each fixed $(N_A,N_B)$ sector, the model is unitarily equivalent to two decoupled hard-core chains described by $H(0)$. To achieve this, we define a non-local unitary
\begin{equation}
\label{eq:Ustat}
U_{\mathrm{stat}}(\theta)
=
\exp\!\left(i\theta\sum_{s<r} n_{r,A} n_{s,B}\right),
\end{equation}
that is able to remove the local statistical phases through conjugation.

For open boundaries, direct conjugation of each bond gives
\begin{equation}
\label{eq:sector_equiv}
H_{\mathrm{dec}}^{\mathrm{OBC}} \equiv U_{\mathrm{stat}}^{\dagger}(\theta)  H(\theta) U_{\mathrm{stat}}(\theta) = H^{\mathrm{hc}}(0),
\end{equation}
where $H_{\mathrm{dec}}^{\mathrm{OBC}}(\theta)$ is the (conjugated) decoupled Hamiltonian which becomes equal to the hardcore bosons Hamiltonian on two disjointed chains, $H^{\mathrm{hc}}(0)$.
Therefore, the local phases are removable, due to the fact that they have opposite signs on the two legs.\footnote{With equal signs, for example, as would be generated by a single anyonic Jordan--Wigner string along the zig-zag, the configuration-space loop in which an $A$ and a $B$ particle on rung $j$ hop in turn to rung $j+1$ and back encloses a flux $2\theta$. No diagonal unitary can remove this flux, and the OBC spectrum then depends on $\theta$.}

Thus OBC energies are exactly independent of $\theta$, but the eigenvectors generally do not, or in other words the eigenstates are ``dressed states,'' as is now described. If $|\Phi\rangle$ is an eigenstate of $H(0)$ with eigenvalue $E$, where $|\Phi\rangle=|\psi_A\rangle\otimes|\psi_B\rangle$ is a product state, the eigenstate of $H(\theta)$ is therefore
\begin{equation}
|\Psi(\theta)\rangle = U_{\mathrm{stat}}(\theta)|\Phi\rangle,
\end{equation}
with the same eigenvalue $E$. 

For PBC the transformation removes the phases within the bulk, but for hopping on a leg around the boundary (i.e. $L\leftrightarrow 1$), the transformation accumulates the phases on the opposite leg. This is shown explicitly in the Appendix~\ref{App A: Mapping to hardcore bosons}~. Therefore, the $H(\theta)$ becomes unitarily equivalent to two decoupled rings, with fluxes fixed by the particle number of the \emph{other} leg. Suppose the Hamiltonian of each ``conjugated'' ring is defined as 
\begin{equation}
H^{\mathrm{hc}}_{\mathrm{ring}}(\varphi)=-t\left[\sum_{j=1}^{L-1}b_{j+1}^{\dagger}b_j
+e^{i\varphi}b_1^{\dagger}b_L+\mathrm{h.c.}\right],
\end{equation}
with free bulk but with a boundary flux of $\varphi$. Therefore, the decoupled Hamiltonian (obtained from the conjugation $H_{\mathrm{dec}}^{\mathrm{PBC}} = U_{\mathrm{stat}}^{\dagger}H(\theta)U_{\mathrm{stat}}$) can be written as,
\begin{equation}
\begin{aligned}
H_{\mathrm{dec}}^{\mathrm{PBC}} \equiv H^{\mathrm{hc}}_{\mathrm{ladder}}(0; \varphi_A, \varphi_B)
&=H_A^{\mathrm{hc}}(\varphi_A)+H_B^{\mathrm{hc}}(\varphi_B),\\
\varphi_A&=\theta N_B,
\qquad \varphi_B=-\theta N_A,
\end{aligned}
\label{eq:hc_fluxes}
\end{equation}
where the boundary fluxes on the two rings are indicated.

After applying a one-dimensional Jordan--Wigner transformation \cite{JordanWigner1928,LiebSchultzMattis1961} to each leg, the corresponding fermionic twists are
\begin{equation}
\begin{aligned}
\Phi_A&=\pi(N_A-1)-\theta N_B,\\
\Phi_B&=\pi(N_B-1)+\theta N_A
\quad (\mathrm{mod}\ 2\pi).
\end{aligned}
\label{eq:fermion_twists}
\end{equation}
Finite-size PBC energies therefore generally depend on $\theta$. The mapped product state also depends on the induced fluxes, $|\Phi_\theta\rangle=|\psi_A(\varphi_A)\rangle\otimes|\psi_B(\varphi_B)\rangle$, with $|\psi_\ell\rangle$ an eigenvector of $H^{\mathrm{hc}}_\ell(\varphi_\ell)$ for $\ell \in \{\mathrm{A}, \mathrm{B}\}$. The eigenvalues of the original Hamiltonian $H(\theta)$ are therefore sums $E_A+E_B$ of single-ring energies, and its eigenvectors are 
\begin{equation}
U_{\mathrm{stat}}(\theta)\,|\psi_A(\varphi_A) \rangle\otimes|\psi_B(\varphi_B) \rangle.
\end{equation}
Therefore, for both open and periodic boundaries, the eigenvectors of $H(\theta)$ are obtained from nonlocal rotations of the tensor product of the eigenvectors of two independent hardcore chains. That is, the eigenvectors of leg A and leg B of the ladder are coupled by the nonlocal unitary $U_{\mathrm{stat}}(\theta)$.

\begin{figure*}[t]
\centering
\begin{tikzpicture}[x=1.6cm,y=1.3cm, every node/.style={font=\small}]
	\foreach \j in {0,...,4}{
		\coordinate (A\j) at (\j,1.4);
		\coordinate (B\j) at (\j,0.0);
	}
	\draw[thick] (A0) -- (A4);
	\draw[thick] (B0) -- (B4);
	\foreach \j in {0,...,4}{
		\draw[gray!50] (A\j) -- (B\j);
		\filldraw[white,draw=black] (A\j) circle (0.07);
		\filldraw[white,draw=black] (B\j) circle (0.07);
	}
	\node[left=0.2cm of A0] {Leg $A$};
	\node[left=0.2cm of B0] {Leg $B$};
	\node[above=0.12cm of A0] {$1$};
	\node[above=0.12cm of A4] {$L$};
	\draw[dashed,thick] (-0.35,0.7) -- (4.35,0.7);
	\node[right] at (-1.5,0.7) {Leg cut};
	\draw[->,thick] (1.1,1.72) -- (1.9,1.72);
	\node[above] at (1.5,1.72) {$e^{+i\theta n_{j,B}}$};
	\draw[->,thick] (1.1,-0.32) -- (1.9,-0.32);
	\node[below] at (1.5,-0.32) {$e^{-i\theta n_{j+1,A}}$};
	\draw[densely dashed,thick] (B1) .. controls +(0.0,0.7) and +(0.0,-0.7) .. (A3);
	\node[anchor=west] at (4.55,1.35) {$U_{\mathrm{stat}}(\theta)$:};
	\node[anchor=west] at (4.55,0.95) {$e^{i\theta\sum_{r>s} n_{r,A}n_{s,B}}$};
	\node[anchor=west] at (4.55,0.5) {ordered pairs $r>s$};
\end{tikzpicture}
\caption{Microscopic setup of the statistical-dressing ladder in the convention of Eq.~\eqref{eq:Hladder}. Hopping is only along the two legs, but each hop carries a density-dependent Peierls phase set by the opposite leg. The dashed horizontal line indicates the leg bipartition, while the dashed interleg curve indicates that $U_{\mathrm{stat}}$ counts ordered $A$--$B$ pairs with $r>s$.}
\label{fig:model}
\end{figure*}

\section{Leg-Cut Reduced Density Matrix}
\label{sec:fcs}
In this section, we describe how to obtain the reduced state of the ladder. Our analysis shall mainly involve only the leg bipartition $[\mathrm{Leg~A}:\mathrm{Leg~B}]$ because the statistical string acts irreducibly across it. For a spatial cut along the rung direction, by contrast, the dressing factorizes into unitaries supported on the two sides of the cut and leaves all Schmidt values invariant for both OBC and PBC. This implies the entanglement spectrum does not ``flow'' with the statistical angle $\theta$. The proof of this is given in Appendix~\ref{App: trivial rung-cut}. We therefore focus on the leg-cut reduced state.

The result is presented in two complementary forms, namely, in terms of a random unitary, completely positive and trace-preserving (CPTP) map and also in terms of an influence functional. We first present the first form, and then follow by the second. From here on we will write for the leg occupational number configurations on leg~A $\mathbf m_A$ as $ \mathbf m$, and that of leg~B  $\mathbf n_B$ as $\mathbf n$, where the leg labels are dropped.  

\subsection{leg-cut reduced state as a random unitary map}
Consider an arbitrary product density operator in the decoupled frame,
\begin{equation}
\sigma_{AB}=\sigma_A\otimes\sigma_B.
\label{eq:product_input}
\end{equation}
The rotated density operator $\rho_{AB}(\theta)$ is
\begin{equation}
\rho_{AB}(\theta)=U_{\mathrm{stat}}(\theta)\sigma_{AB}U_{\mathrm{stat}}^{\dagger}(\theta).
\label{eq:rotated_density}
\end{equation}

Without any special importance attached to any leg, we shall obtain the reduced state of leg~A. To that end, let $|\mathbf{n} \rangle = |n_1, n_2, \ldots, n_L\rangle$ denote an occupational configuration on leg~B, and the occupational probability on leg~B
\begin{equation}
p_n =\langle \mathbf n |\sigma_B| \mathbf n \rangle,
\end{equation}
the suffix field operator
\begin{equation}
B^{>}_{s,A} =\sum_{r>s}n_{r,A},
\end{equation}
and a unitary operator obtained for leg~A,
\begin{equation}
V_{\mathbf n}(\theta) =\exp\!\left(i\theta\sum_s n_s{(\mathbf n)}B^{>}_{s,A}\right),
\label{eq:controlled_unitaries}
\end{equation} 
after conditioning on the leg~B occupational configuration $\mathbf n$, and the $n_s{(\mathbf n)}$ also is the particle number for site $s$ in the same configuration ($|\mathbf n \rangle$). Since $U_{\mathrm{stat}}$ is diagonal in this basis,
\begin{equation}
U_{\mathrm{stat}}(\theta)=\sum_n V_{\mathbf n}(\theta) \otimes | \mathbf n\rangle\langle \mathbf n|,
\end{equation}
written as a controlled operation conditioned on the leg~B occupational number states $| \mathbf n\rangle$.

Taking the partial trace of $\rho_{AB}(\theta)$ gives
\begin{equation}
\begin{aligned}
\rho_A(\theta)&=\mathcal E_\theta(\sigma_A)\\
&=\sum_{\mathbf n} p_{\mathbf n} V_{\mathbf n}(\theta) \sigma_A V_{\mathbf n}^{\dagger}(\theta).
\end{aligned}
\label{eq:rho_exact_fcs}
\end{equation}
This expression is a random-unitary, completely positive, trace-preserving, and unital map \cite{AudenaertScheel2008}. 
The expression means leg~B acts on leg~A as classical randomness with probability $p_{\mathbf n}$, where leg~A is rotated by $V_{\mathbf n}(\theta)$; The reduced state $\rho_A(\theta)$ only depends on leg~B through the occupation probabilities $p_n$ of $\sigma_B$. 

Similarly, if leg~A was rather traced out, the reduced state of leg~B would keep all of $\sigma_B$ but now weighted by the occupational probabilities of leg~A. For the occupational number states on leg~A, $\mathbf m$, the reduced state on leg~B is 
\begin{equation}
\rho_B=\sum_{\mathbf m} q_{\mathbf m} W_{\mathbf m}(\theta) \sigma_B W_{\mathbf m}^\dagger (\theta),
\end{equation}
which depends on $\sigma_A$ only through $q_{\mathbf m}=\langle\mathbf m |\sigma_A|\mathbf m \rangle$, and $W_{\mathbf m}(\theta)$ is defined similarly to $V_{\mathbf n}(\theta)$.

The bound for the leg-cut entanglement entropy of $S(\rho_A)$ proceed from the expressions above. For a general product input, concavity and the entropy bound for a classical mixture \cite{NielsenChuang2010} imply
\begin{equation}
\begin{aligned}
S(\sigma_A)&\le S\!\left(\rho_A(\theta)\right)
\le S(\sigma_A)+H(p),\\
H(p)&=-\sum_{\mathbf n}p_{\mathbf n}\log p_{\mathbf n}.
\end{aligned}
\label{eq:entropy_bounds}
\end{equation}
For a pure input, $0\le S(\rho_A)\le H(p)$, and the rank of $\rho_A$ is bounded by the number of configurations of leg~B that occur,
\begin{equation}
\operatorname{rank}\rho_A(\theta)\le\#\{\mathbf n:p_{\mathbf n}>0\}.
\end{equation}

\subsection{leg-cut reduced state as an influence functional} 
To realize the leg-cut reduced state as an influence functional, we start with the fact that every $V_{\mathbf n}(\theta)$ is a function of the commuting suffix charges $B^{>}_{s,A}$, and is therefore diagonal in the occupation basis $\{|\mathbf m\rangle_A\}$ of leg~A. Let $b_s(\mathbf m)=\sum_{r>s}m_r$ be the eigenvalue of $B^{>}_{s,A}$ on $|\mathbf m\rangle_A$, and
\begin{equation}
\beta_s(\mathbf m,\mathbf m')=b_s(\mathbf m)-b_s(\mathbf m')=\sum_{r>s}\left(m_r-m'_r\right)
\label{eq:beta_def}
\end{equation}
the difference of the suffix charges of two retained-leg configurations. Then $V_{\mathbf n}(\theta)|\mathbf m\rangle_A=z_{\mathbf m}(\mathbf n)|\mathbf m\rangle_A$ with the phase $z_{\mathbf m}(\mathbf n)=\exp\!\big(i\theta\sum_sn_sb_s(\mathbf m)\big)$, and the matrix elements of Eq.~\eqref{eq:rho_exact_fcs} factorize,
\begin{equation}
\begin{aligned}
\rho^A_{\mathbf m\mathbf m'}(\theta)&=\sigma^A_{\mathbf m\mathbf m'}F_{\mathbf m\mathbf m'}(\theta),\\
F_{\mathbf m\mathbf m'}(\theta)&=\sum_{\mathbf n}p_{\mathbf n}\,z_{\mathbf m}(\mathbf n)\,z_{\mathbf m'}(\mathbf n)^*.
\end{aligned}
\label{eq:F_gram}
\end{equation}
Since $z_{\mathbf m}(\mathbf n)z_{\mathbf m'}(\mathbf n)^*=\exp\!\big(i\theta\sum_sn_s\beta_s(\mathbf m,\mathbf m')\big)$ is the exponential of a linear function of the occupations of leg~B, the sum over $\mathbf n$ is an expectation value in $\sigma_B$,
\begin{equation}
\begin{aligned}
F_{\mathbf m\mathbf m'}(\theta)&=\chi_B\!\left[\theta\boldsymbol\beta(\mathbf m,\mathbf m')\right],\\
\chi_B[\boldsymbol\lambda]&\equiv\operatorname{Tr}_B\!\left[\sigma_B\exp\!\Big(i\sum_s\lambda_sn_{s,B}\Big)\right].
\end{aligned}
\label{eq:fcs_factor}
\end{equation}
Here $\chi_B[\boldsymbol\lambda]=\sum_{\mathbf n}p_{\mathbf n}e^{i\boldsymbol\lambda\cdot\mathbf n}$ is the characteristic function of the occupation distribution of leg~B, that is, the generating function of its full counting statistics (FCS) \cite{Levitov1996,Klich2003}, with one counting field $\lambda_s$ per rung. The counting fields are not auxiliary probes: they are fixed by the microscopic statistical string, $\lambda_s=\theta\beta_s(\mathbf m,\mathbf m')$. The positive sign follows from the convention $|\Psi(\theta)\rangle=U_{\mathrm{stat}}(\theta)|\Phi\rangle$.

This factor is the influence functional of leg~B on leg~A. To see why, write the string generator as $G=\sum_sB^{>}_{s,A}\,n_{s,B}$, so that $U_{\mathrm{stat}}$ is also a controlled operation in the opposite direction,
\begin{equation}
\begin{aligned}
U_{\mathrm{stat}}(\theta)&=\sum_{\mathbf m}|\mathbf m\rangle\langle\mathbf m|_A\otimes W_{\mathbf m}(\theta),\\
W_{\mathbf m}(\theta)&=\exp\!\Big(i\theta\sum_sb_s(\mathbf m)\,n_{s,B}\Big),
\end{aligned}
\label{eq:W_controlled}
\end{equation}
in which the configuration of leg~A selects a unitary on leg~B. Taking the matrix element of $\rho_A=\operatorname{Tr}_B[U_{\mathrm{stat}}(\sigma_A\otimes\sigma_B)U_{\mathrm{stat}}^\dagger]$ in this basis gives directly
\begin{equation}
F_{\mathbf m\mathbf m'}(\theta)=\operatorname{Tr}_B\!\left[W_{\mathbf m}(\theta)\,\sigma_B\,W_{\mathbf m'}^{\dagger}(\theta)\right],
\label{eq:F_overlap}
\end{equation}
the overlap of the states of the traced leg conditioned on the two retained-leg configurations. This is the structure of a Feynman--Vernon influence functional \cite{FeynmanVernon1963}: leg~A is the system, leg~B is the environment, $\mathbf m$ and $\mathbf m'$ label the system on the forward and backward branches, and $F$ is the response of the environment to that pair. Because the coupling is diagonal in the occupation basis of both legs, there is no sum over histories: a single pair $(\mathbf m,\mathbf m')$ suffices, and the influence functional reduces to the characteristic function $\chi_B[\boldsymbol\lambda]$ of Eq.~\eqref{eq:fcs_factor} at the counting fields generated by that pair. We call the matrix $F_{\mathbf m\mathbf m'}$ the influence matrix, which is also a Gram matrix. Consequently,
\begin{equation}
\begin{aligned}
F_\theta&\succeq0,
&F_{\mathbf m, \mathbf m} &=1,\\
F_{\mathbf {m'}, \mathbf m} &=F_{\mathbf m, \mathbf {m'}}^*,
&|F_{\mathbf m, \mathbf {m'}}|&\le1.
\end{aligned}
\end{equation}
Thus $\rho_A=\sigma_A\circ F_\theta$ is a Schur multiplier by a correlation matrix. The Schur product theorem \cite{HornJohnson2012} independently establishes positivity, while the unit diagonal establishes trace preservation. The channel preserves all retained-leg occupation probabilities and modifies only their coherences. Under PBC this comparison is made with the decoupled input at the same boundary flux.

Because $n_{s,B}$ is invariant under the Jordan--Wigner map, Eq.~\eqref{eq:fcs_factor} can be evaluated with the fermionic image of $\sigma_B$. If that image is a number-conserving Gaussian state, as for a hard-core-chain ground state, with one-body correlator $(C_B)_{st}=\operatorname{Tr}(\sigma_Bc_s^\dagger c_t)$, the Klich determinant gives \cite{Klich2003,Schonhammer2007}
\begin{equation}
\label{eq:det_form}
\begin{aligned}
F_{\mathbf m, \mathbf {m'}}(\theta)
&=\det\!\left(I-C_B+e^{i\Lambda_{\mathbf m, \mathbf {m'}}}C_B\right),\\
\Lambda_{\mathbf m, \mathbf {m'}}
&=\operatorname{diag}\!\left((\lambda_1)_{\mathbf m, \mathbf {m'}},\ldots,
(\lambda_L)_{\mathbf m, \mathbf {m'}}\right).
\end{aligned}
\end{equation}
This reduces the traced-leg many-body sum to an $L\times L$ determinant. Gaussianity of $\sigma_B$ does not imply Gaussianity of $\rho_A$: occupation-number FCS of a Slater determinant generally has connected cumulants at all orders \cite{KlichLevitov2009,Song2012}, and Eq.~\eqref{eq:rho_exact_fcs} is generally a mixture of distinct Gaussian states rather than one Gaussian state \cite{Peschel2003,PeschelEisler2009,CasiniHuerta2009}.

We end this section with an important note that, the results derived above only applies to a pure product eigenstate and a maximally mixed projector on a factorized degenerate product space. For an entangled input, such as a coherent superposition across a degenerate product manifold, the formula does not apply.

\section{Exact Cumulant Superoperator Hierarchy}
\label{sec:cumulants}
Section~\ref{sec:fcs} showed that leg~B acts on leg~A only through the characteristic function of its occupations. We now expand its logarithm in connected cumulants, and writes the channel $\mathcal E_\theta$ as the exponential of commuting superoperators (linear maps on operators), and we state when this expansion holds.

The characteristic function, rather than its logarithm, is the globally exact object. Recall the characteristic function $\chi_B[\lambda]$ of Eq.~\ref{eq:fcs_factor}, with $\chi_B[0]=1$. There is therefore a zero-free neighborhood of the origin in which a continuous logarithm
\begin{equation}
W_B[\lambda] \equiv \log \chi_B[\lambda]
\end{equation}
is analytic. Because the occupations commute, its Taylor expansion in that neighborhood is the ordinary linked-cumulant expansion
\begin{equation}
\label{eq:W_expand}
W_B[\lambda]
=
\sum_{p=1}^{\infty}
\frac{i^p}{p!}
\sum_{s_1,\dots,s_p}
\lambda_{s_1}\cdots\lambda_{s_p}
\kappa^{(p)}_{s_1\dots s_p},
\end{equation}
with connected density cumulants
\begin{equation}
\kappa^{(p)}_{s_1\dots s_p}
=\langle n_{s_1,B}\cdots n_{s_p,B}\rangle_c.
\end{equation}
This series converges only within the analytic domain bounded by zeros of $\chi_B$ in the relevant complex counting-field direction. Outside it, Eq.~\eqref{eq:fcs_factor} remains exact, but a single logarithmic branch and its Taylor series need not.

Let $\operatorname{ad}_X(Y)=[X,Y]$. Since the suffix charges commute, the exact channel can be written globally without taking a logarithm as
\begin{equation}
\label{eq:superoperator_exact}
\mathcal E_\theta
=\chi_B\!\left(\theta\operatorname{ad}_{\boldsymbol B_A^>}\right).
\end{equation}
Within the analytic domain of Eq.~\eqref{eq:W_expand}, this becomes
\begin{equation}
\rho_A(\theta)=e^{\mathcal L_\theta}\sigma_A,
\end{equation}
where
\begin{equation}
\label{eq:superoperator_def}
\mathcal L_\theta
=
\sum_{p=1}^{\infty}
\frac{(i\theta)^p}{p!}
\sum_{s_1,\dots,s_p}
\kappa^{(p)}_{s_1\dots s_p}
\operatorname{ad}_{B^{>}_{s_1,A}}\cdots \operatorname{ad}_{B^{>}_{s_p,A}}.
\end{equation}
The adjoint actions commute, so there is no Baker--Campbell--Hausdorff ordering problem. The restriction is analytic convergence, not operator ordering.

It follows that, in every zero-free neighborhood where $W_B=\log\chi_B$ is analytic, the FCS channel admits the commuting expansion in Eq.~\eqref{eq:superoperator_def}. Its $p$th term is generated by the connected $p$-point occupation cumulant of the traced leg, whereas the globally exact content is carried by Eqs.~\eqref{eq:rho_exact_fcs} and \eqref{eq:superoperator_exact}.

The hierarchy may still be resolved into local retained-leg densities,
\begin{equation}
B^{>}_{s,A} = \sum_r H_{sr} n_{r,A},
\qquad
H_{sr}=
\begin{cases}
1, & r>s,\\
0, & r\le s.
\end{cases}
\end{equation}
This induces transformed kernels
\begin{equation}
\mu_r = \sum_s H_{sr}\kappa^{(1)}_s,
\end{equation}
\begin{equation}
K^{(2)}_{rr'} = \sum_{s,t} H_{sr}\kappa^{(2)}_{st}H_{tr'},
\end{equation}
\begin{equation}
K^{(3)}_{rr'r''} = \sum_{s,t,u} H_{sr}\kappa^{(3)}_{stu}H_{tr'}H_{ur''},
\end{equation}
and analogously at all higher orders. The first few terms in the hierarchy are therefore
\begin{align}
\mathcal L_1 &= +i\theta \sum_r \mu_r\operatorname{ad}_{n_{r,A}},\\
\mathcal L_2 &= -\frac{\theta^2}{2}\sum_{r,r'} K^{(2)}_{rr'}\operatorname{ad}_{n_{r,A}}\operatorname{ad}_{n_{r',A}},\\
\mathcal L_3 &= -\frac{i\theta^3}{6}\sum_{r,r',r''} K^{(3)}_{rr'r''}\operatorname{ad}_{n_{r,A}}\operatorname{ad}_{n_{r',A}}\operatorname{ad}_{n_{r'',A}},
\end{align}
with higher density-difference superoperators generated for $p\ge4$. These are terms in a channel acting on $\sigma_A$.

\section{Channel Interpretation and the First Entangling Correction}
\label{sec:modular}
We now give the first two cumulants of the cumulant superoperator hierarchy a physical meaning, namely a unitary rotation and correlated dephasing, respectively. We use the exact channel to obtain the first entangling correction, and the loss of purity of a pure product input.

Let
\begin{equation}
\overline n_s=\operatorname{Tr}(\sigma_Bn_{s,B}),
\qquad
M_1=\sum_s\overline n_sB^{>}_{s,A}.
\end{equation}
The first cumulant is
\begin{equation}
\mathcal L_1 = i\theta\operatorname{ad}_{M_1},
\end{equation}
and therefore
\begin{equation}
e^{\mathcal L_1}(X)=e^{i\theta M_1}Xe^{-i\theta M_1}.
\end{equation}
It is a deterministic one-body phase rotation and preserves every eigenvalue and the purity of the input state. For a pure product input it cannot generate leg entanglement.

Define the traced-leg covariance matrix
\begin{equation}
(\kappa_B)_{st}=\langle n_{s,B}n_{t,B}\rangle
-\langle n_{s,B}\rangle\langle n_{t,B}\rangle.
\end{equation}
The second cumulant is the correlated-dephasing generator
\begin{equation}
\mathcal L_2=-\frac{\theta^2}{2}\sum_{s,t}(\kappa_B)_{st}
\operatorname{ad}_{B^{>}_{s,A}}\operatorname{ad}_{B^{>}_{t,A}}.
\end{equation}
Since $\kappa_B\succeq0$, let $\boldsymbol\xi$ be a real Gaussian vector with $\mathbb E(\xi_s\xi_t)=\theta^2(\kappa_B)_{st}$. Then
\begin{equation}
e^{\mathcal L_1+\mathcal L_2}(X)
=\mathbb E_{\boldsymbol\xi}\!\left[
e^{i\sum_s(\theta\overline n_s+\xi_s)B^{>}_{s,A}}
X
e^{-i\sum_s(\theta\overline n_s+\xi_s)B^{>}_{s,A}}
\right].
\label{eq:gaussian_noise_channel}
\end{equation}
Thus the first-two-cumulant approximation is itself a completely positive, trace-preserving, unital random-unitary channel.

Within this local hierarchy, the first cumulant is unitary, while the second is the first term that can generate mixedness and a nontrivial entanglement spectrum from a pure product input. However, it should be noted that a finite exponential truncation containing third or higher cumulants need not be positive \cite{Marcinkiewicz1939,RajagopalSudarshan1974}. Such a truncation cannot be interpreted as a density matrix, or assigned an entanglement spectrum, without a positivity test. In particular, eigenvalue clipping would constitute an additional nonlinear approximation rather than a property of the cumulant expansion.

The exact channel also yields a direct entanglement formula. For a pure product input $|\psi_A\rangle=\sum_{\mathbf m }\psi_{\mathbf m} | \mathbf m\rangle$, Eq.~\eqref{eq:rho_exact_fcs} gives
\begin{equation}
\operatorname{Tr}\rho_A(\theta)^2
=\sum_{\mathbf m,\mathbf {m'}}|\psi_{\mathbf m} |^2|\psi_{\mathbf {m'}}|^2
\left|\chi_B\!\left(\theta\boldsymbol\beta(\mathbf m, \mathbf {m'})\right)\right|^2.
\label{eq:exact_purity}
\end{equation}
Because $|\chi_B|\le1$, this purity cannot exceed one. Define the retained suffix-charge covariance
\begin{equation}
(\Gamma_A)_{st}
=\langle B^{>}_{s,A}B^{>}_{t,A}\rangle_A
-\langle B^{>}_{s,A}\rangle_A\langle B^{>}_{t,A}\rangle_A.
\end{equation}
Expanding the squared characteristic function and averaging over two independent retained-leg configurations yields
\begin{equation}
1-\operatorname{Tr}\rho_A(\theta)^2
=2\theta^2\operatorname{Tr}(\kappa_B\Gamma_A)+O(\theta^4).
\label{eq:purity_coefficient}
\end{equation}
Both covariance matrices are positive semidefinite, so $\operatorname{Tr}(\kappa_B\Gamma_A)\ge0$. The leading entanglement is generated precisely when traced-leg occupation fluctuations overlap with retained-leg suffix-charge fluctuations. 

The second R\'enyi entropy obeys
\begin{equation}
S_2(\theta)=2\theta^2\operatorname{Tr}(\kappa_B\Gamma_A)+O(\theta^4).
\end{equation}
In general, there is no known or simple closed-form expression for the entanglement entropy, whether von Neumann or R\'enyi, to the best of our knowledge. The origin of this difficulty is related to that of the entanglement Hamiltonian $-\log \rho$. Such an expansion is singular about a rank-one product state outside its support. 

\section{Numerical Validation Results}
\label{sec:numerics}

We now numerically test the exact identities of the previous theory sections, namely Secs.~\ref{sec:model}, Sec~\ref{sec:fcs}, the first-two-cumulant approximation of Sec.~\ref{sec:modular}, and the purity coefficient of Eq.~\eqref{eq:purity_coefficient}, by exact diagonalization of the ladder in fixed particle-number sectors, for both boundary conditions up to a ladder size of $L\le8$. In each case we construct the nondegenerate ground state $|\Phi\rangle$ of the decoupled chains $H_{\mathrm{dec}}$, which is $H(0)$ for OBC and $H'_{\mathrm{PBC}}(\theta)$ for the the pair of hard-core rings with the fluxes as given in Eq.~\eqref{eq:hc_fluxes} for PBC, and dress it, $|\Psi\rangle=U_{\mathrm{stat}}|\Phi\rangle$.

We then compare the following: (i) $H(\theta)$ with $U_{\mathrm{stat}}H_{\mathrm{dec}}U_{\mathrm{stat}}^\dagger$; (ii) $H(\theta)|\Psi\rangle$ with $E|\Psi\rangle$, where $E$ is the sum of the ground-state energies of the two decoupled chains; (iii) the direct partial trace of $|\Psi\rangle\langle\Psi|$ over leg~B with Eq.~\eqref{eq:rho_exact_fcs}; and (iv) the direct FCS sum with the determinant of Eq.~\eqref{eq:det_form}. Matrix errors are relative Frobenius norms, and the eigenstate residual $\|H\Psi-E\Psi\|$ is normalized by $\|H\Psi\|$. In every case, (i)--(iv) hold to relative error $\lesssim2\times10^{-15}$. By (i), $|\Psi\rangle$ is the ground state of $H(\theta)$, so $\rho_A$ is the leg-cut reduced state of the physical ground state.

The first-two-cumulant approximation is $\rho_A^{[2]}=e^{\mathcal L_1+\mathcal L_2}(\sigma_A)$. It keeps the mean and covariance of the traced-leg occupations and drops all higher connected cumulants. We test the validity of this approximation by computing two quantities: (i)~the trace distance, $D_{\mathrm{tr}}=\frac12\|\rho_A-\rho_A^{[2]}\|_1$, of the approximate state from the exact reduced state, which gives the largest probability difference any measurement on leg A can see,  and (ii)~the smallest eigenvalue $\lambda_{\min}^{[2]}$ of $\rho_A^{[2]}$, which tests positivity. Table~\ref{tab:validation} reports these values for representative nondegenerate cases, from $(N_A,N_B)=(1,1)$ at $L=4$ to $(3,3)$ at $L=8$.

\begin{table}[t]
\caption{First-two-cumulant approximation. $D_{\mathrm{tr}}$ is the trace distance between the exact reduced state $\rho_A$ of leg~A and the approximation $\rho_A^{[2]}$, and $\lambda_{\min}^{[2]}$ is the smallest eigenvalue of $\rho_A^{[2]}$.}
\label{tab:validation}
\begin{ruledtabular}
\begin{tabular}{cccccc}
BC & $L$ & $(N_A,N_B)$ & $\theta$ & $D_{\mathrm{tr}}$ & $\lambda_{\min}^{[2]}$\\
\hline
OBC & 4 & $(1,1)$ & 0.370 & $3.63\times10^{-4}$ & $9.13\times10^{-4}$\\
OBC & 5 & $(2,1)$ & 0.710 & $4.62\times10^{-3}$ & $2.54\times10^{-5}$\\
OBC & 5 & $(2,3)$ & 0.710 & $3.83\times10^{-3}$ & $5.49\times10^{-5}$\\
OBC & 6 & $(2,1)$ & 1.130 & $2.12\times10^{-2}$ & $1.89\times10^{-5}$\\
OBC & 6 & $(2,2)$ & 1.130 & $2.07\times10^{-2}$ & $6.44\times10^{-5}$\\
OBC & 6 & $(3,3)$ & 1.130 & $1.21\times10^{-2}$ & $7.87\times10^{-6}$\\
OBC & 7 & $(3,3)$ & 1.130 & $2.08\times10^{-2}$ & $8.95\times10^{-7}$\\
OBC & 8 & $(3,3)$ & 1.130 & $2.60\times10^{-2}$ & $8.24\times10^{-8}$\\
PBC & 4 & $(1,1)$ & 0.370 & $4.18\times10^{-4}$ & $2.13\times10^{-3}$\\
PBC & 5 & $(2,1)$ & 0.710 & $5.13\times10^{-3}$ & $1.38\times10^{-4}$\\
PBC & 5 & $(2,3)$ & 0.710 & $3.96\times10^{-3}$ & $2.74\times10^{-4}$\\
PBC & 6 & $(2,1)$ & 1.130 & $2.37\times10^{-2}$ & $2.67\times10^{-4}$\\
PBC & 6 & $(2,2)$ & 1.130 & $2.15\times10^{-2}$ & $7.10\times10^{-4}$\\
PBC & 6 & $(3,3)$ & 1.130 & $1.32\times10^{-2}$ & $9.54\times10^{-5}$\\
PBC & 7 & $(3,3)$ & 1.130 & $2.04\times10^{-2}$ & $3.79\times10^{-5}$\\
PBC & 8 & $(3,3)$ & 1.130 & $2.64\times10^{-2}$ & $1.01\times10^{-5}$\\
\end{tabular}
\end{ruledtabular}
\end{table}

The first-two-cumulant output is positive in every case, as guaranteed by Eq.~\eqref{eq:gaussian_noise_channel}, and $\lambda_{\min}^{[2]}\ge8\times10^{-8}$ is far above rounding error. (This eigenvalue belongs to the approximation: the exact $\rho_A$ can have zero eigenvalues, since its rank is at most the number of leg-$B$ configurations.) The approximation is accurate, $D_{\mathrm{tr}}<3\times10^{-2}$, and $D_{\mathrm{tr}}$ grows from about $4\times10^{-4}$ at $\theta=0.37$ to $(1$--$3)\times10^{-2}$ at $\theta=1.13$, although the table also increases $L$ along this sequence. At fixed $\theta=1.13$, $D_{\mathrm{tr}}$ for $(3,3)$ grows with $L$. We found that at small $\theta$ the distance scales as $\theta^{3}$, as expected since the leading omitted term involves the third cumulant $\kappa^{(3)}$ of the traced-leg densities. But for a half-filled traced leg ($N_B=L/2$), particle--hole symmetry forces $\kappa^{(3)}=0$ and the scaling is $\theta^{4}$. The cases shown in the table are only illustrative, so they do not necessarily prove  uniform convergence in the thermodynamic limit.

We also test the purity coefficient of the two sides of Eq.~\eqref{eq:purity_coefficient} for OBC. For all cases where the approximation holds, i.e. where $\theta$ is small, we obtain their relative difference to about $(1 - 2)\times 10^{-7}$. For example, for a ladder with $L=6$, $(N_A,N_B)=(2,1)$, and $\theta=10^{-3}$, direct evaluation gives
\begin{equation}
\frac{1-\operatorname{Tr}\rho_A(\theta)^2}{\theta^2}
=0.19905345,
\end{equation}
whereas the covariance prediction is
\begin{equation}
2\operatorname{Tr}(\kappa_B\Gamma_A)=0.19905347.
\end{equation}
Their relative difference of about $10^{-7}$ is consistent with the omitted $O(\theta^4)$ term in the purity, a relative correction of order $\theta^2$. The same test for $(N_A,N_B)=(2,2)$ and $(3,3)$ at $L=6$ and $(2,3)$ at $L=5$ yields relative differences of about $(1$--$2)\times10^{-7}$.

\section{Discussion}
\label{sec:discussion}

The leg-cut reduced state is exactly solvable because $U_{\mathrm{stat}}$ is diagonal in the occupation basis of the traced leg. Tracing out leg~B replaces the controlled unitaries $V_{\mathbf n}(\theta)$ by a classical average over the occupation configuration $\mathbf n$ of leg~B, Eq.~\eqref{eq:rho_exact_fcs}. Three facts follow: the map is completely positive, trace preserving, and unital. It depends on leg~B only through the occupation probabilities $p_{\mathbf n}$, so coherences of $\sigma_B$ play no role. And for a pure product input the leg-cut entanglement entropy cannot exceed the Shannon entropy $H(p)$ of the occupation distribution of the traced leg, Eq.~\eqref{eq:entropy_bounds}. This ceiling is independent of $\theta$, so it does not estimate the entropy at small $\theta$. In the occupation basis of leg~A, the channel multiplies $\sigma_A$ elementwise by the influence matrix of Eq.~\eqref{eq:fcs_factor}. Its unit diagonal leaves every occupation probability of leg~A unchanged, so only coherences are affected.

The cumulant hierarchy is a local expansion of the exact channel $\mathcal E_\theta=\chi_B(\theta\,\mathrm{ad}_{\boldsymbol B^{>}_A})$, Eq.~\eqref{eq:superoperator_exact}. Its series, Eq.~\eqref{eq:superoperator_def}, converges only in the zero-free neighborhood of zero counting field where $\log\chi_B$ is analytic. There, the first cumulant is a deterministic phase rotation and the second is correlated dephasing, an average over Gaussian phase noise with kernel $\kappa_B\succeq0$, Eq.~\eqref{eq:gaussian_noise_channel}, so the second-order channel is completely positive. Truncations at third or higher order carry no such guarantee, and they need a positivity test before an entanglement spectrum is assigned to them. In Table~\ref{tab:validation} ($L\le8$, $\theta\le1.13$), the second-order state is within $D_{\mathrm{tr}}<3\times10^{-2}$ of the exact one, but $D_{\mathrm{tr}}$ grows with $L$ at fixed $\theta$, so convergence in the thermodynamic limit is not established.

Boundary conditions determine the spectrum and the product state $|\Phi\rangle$, but $\theta$ enters the entanglement only through $U_{\mathrm{stat}}$. Under OBC, $|\Phi\rangle$ is independent of $\theta$. Under PBC the induced fluxes shift the energies and make $|\Phi_\theta\rangle$ depend on $\theta$. Away from the ground-state level crossings, where the first occurs at $|\theta|=\pi/\max(N_A,N_B)$, $|\Phi_\theta\rangle$ differs from the zero-flux state $|\Phi_0\rangle$ only by the on-site phases (shown in the appendix, Eq.~\eqref{eq:flux_phase}). These phases commute with $U_{\mathrm{stat}}$ and leave the occupation probabilities unchanged. Since they act on each leg separately, they leave the leg-cut Schmidt values unchanged. The entanglement spectrum is therefore that of $U_{\mathrm{stat}}(\theta)|\Phi_0\rangle$ for both boundary conditions. At a crossing, the ground state is degenerate, and the formulas apply only to a product state chosen in the degenerate manifold.

For a pure product input, the leading purity loss, Eq.~\eqref{eq:purity_coefficient}, is proportional to the overlap $\operatorname{Tr}(\kappa_B\Gamma_A)$ of two positive semidefinite covariance matrices. The overlap is nonnegative, and it vanishes if and only if the supports of $\kappa_B$ and $\Gamma_A$ are orthogonal, for example when at every rung either the density of leg~B or the suffix charge of leg~A does not fluctuate, since the string couples the two rung-by-rung. Because $\operatorname{Tr}\rho_A^2$ is a finite trigonometric polynomial in $\theta$, Eq.~\eqref{eq:exact_purity}, the purity loss and $S_2$ are analytic at $\theta=0$. The entanglement Hamiltonian $-\log\rho_A$ is not, because $\rho_A(0)$ has rank one and the logarithm diverges outside its support. To our knowledge, the von Neumann entropy has no closed form.

\section{Conclusion}
\label{sec:conclusion}

The central exact statement of this work is that the leg-cut reduced density matrix of the statistical-dressing ladder is an FCS influence functional whose counting fields are fixed by the microscopic string. For every product input it defines an exact random-unitary channel on the retained leg, weighted by the traced-leg occupation probabilities, which preserves the retained-leg occupation probabilities. For a Slater-determinant traced leg its influence-matrix elements are $L\times L$ determinants, although the reduced state is generally not Gaussian. Under periodic boundaries the induced fluxes add only leg-local phases to a nondegenerate ground state, so the entanglement again depends on $\theta$ only through the string. The entropy generated by the string cannot exceed the Shannon entropy of the traced-leg occupation distribution. 

The cumulant expansion of the influence matrix that builds an approximation of the leg-cut reduced state is local: its first term is a unitary rotation, its second a completely positive correlated dephasing, and truncations at third or higher order need not be positive. For a pure product input, the purity loss and the second R\'enyi entropy are, to leading order, $2\theta^2\operatorname{Tr}(\kappa_B\Gamma_A)$, the overlap of the traced-leg occupation covariance and the retained-leg suffix-charge covariance. 

The accuracy of our approximation in the thermodynamic limit, and a closed form for the von Neumann entropy, remain open.

In conclusion, our results establish an exact lattice connection between statistical strings, counting statistics, and entanglement generation.

\section*{Acknowledgements}
The author thanks former collaborators for discussions on works related to ladder models of tight-binding anyons.

\paragraph*{Author contributions:} The sole author conceived the project, carried out the analytical derivations and numerical analysis, prepared the figures, and wrote the manuscript.

\begin{appendix}

\section{Exact mapping of Hamiltonian to hardcore boson chains}
\label{App A: Mapping to hardcore bosons}

This appendix section records the details behind the statements contained in Section~\ref{sec:model} concerning the mapping of the model Hamiltonian to decoupled ordinary hardcore bosonic chains (for OBC) or rings (for PBC). We also discuss the spectrum and eigenstates for both cases.

The Hamiltonian under consideration (as stated in Eq.~\ref{eq:Hladder}) is 
\begin{align}
\label{eq:Hladder_app}
H(\theta) &= -t\sum_j \left( b_{j+1,A}^{\dagger}  e^{+i\theta n_{j,B}} b_{j,A}  \right. \nonumber \\
& \qquad \qquad \left. + b_{j+1,B}^{\dagger} e^{-i\theta n_{j+1,A}} b_{j,B} + \text{h.c.} \right) ,
\end{align}
where for OBC $j$ runs over $\{1, \ldots, L-1 \}$, and for PBC $j$ runs over $ \{1, \ldots, L\}$, in addition with the extra condition that $b_{L+1, \ell} \equiv b_{1, \ell}$, for each leg $\ell \in \{\mathrm{A, B} \}$.

With
\begin{equation}
G=\sum_{r>s}n_{r,A}n_{s,B},
\qquad
U_{\mathrm{stat}}=e^{i\theta G},
\end{equation}
the required commutators are
\begin{align}
&[n_{j,\ell}, b_{j,\ell}] = - b_{j,\ell}, \quad \ell \in \{A,B \}\\
&[G,b_{j,A}] = -b_{j,A}\sum_{s<j}n_{s,B},\\
&[G,b_{j,B}] = -b_{j,B}\sum_{r>j}n_{r,A}.
\end{align}
All number operators in those sums commute with $G$, and so the conjugations exponentiate directly. For any $X$, 
\begin{equation}
e^{i\theta G} X e^{-i\theta G} = e^{i \theta \mathrm{ad_G}} X, \qquad \text{where $\mathrm{ad_G} = [G, \cdot]$ } 
\end{equation}

Therefore,
\begin{align}
U_{\mathrm{stat}}b_{j,A}U_{\mathrm{stat}}^\dagger
&=b_{j,A}e^{-i\theta\sum_{s<j}n_{s,B}},\\
U_{\mathrm{stat}}b_{j,B}U_{\mathrm{stat}}^\dagger
&=b_{j,B}e^{-i\theta\sum_{r>j}n_{r,A}}.
\end{align}
The inverse conjugation flips the sign of the exponent
\begin{align}
U^\dagger b_{j,A}U=b_{j,A}\,e^{+i\theta\sum_{s<j}n_{s,B}},  \\
U^\dagger b_{j,B}U=b_{j,B}\,e^{+i\theta\sum_{r>j}n_{r,A}}.
\end{align}
Consequently, every OBC bond transforms as
\begin{align}
&U_{\mathrm{stat}}^\dagger \left( b_{j+1,A}^\dagger e^{+i\theta n_{j,B}} b_{j,A} \right) U_{\mathrm{stat}}
=b_{j+1,A}^\dagger b_{j,A},\\
&U_{\mathrm{stat}}^\dagger \left( b_{j+1,B}^\dagger e^{-i\theta n_{j+1,A}} b_{j,B} \right) U_{\mathrm{stat}}
=b_{j+1,B}^\dagger b_{j,B}.
\end{align}
The conjugation therefore removes the local density-dependent phases bond by bond, and hence
\begin{equation}
\label{eq:sector_equiv_app}
H'(\theta) \equiv U_{\mathrm{stat}}^{\dagger}(\theta)  H(\theta) U_{\mathrm{stat}}(\theta) = H(0),
\end{equation}
which proves the equivalence relation in Eq.~\eqref{eq:sector_equiv}. Therefore, under the open boundary condition, the Hamiltonian $H(\theta)$ has the same spectrum as $H(0)$, which is that of a two disjointed hardcore-boson chains.

For PBC, the transformations within the bulk also hold true, except for the bonds that wrap around the boundary, connecting sites $1$ and $L$. With the periodic condition, the OBC Hamiltonian gets additional terms for hopping across the boundary, namely 
\begin{equation}
\label{Eq:boundary_termA}
b_{1,A}^\dagger e^{+i\theta n_{L,B}}b_{L,A} + \mathrm{h.c.} \quad\text{on leg A}
\end{equation}
and 
\begin{equation}
\label{Eq:boundary_termB}
b_{1,B}^\dagger e^{-i\theta n_{1,A}}b_{L,B} + \mathrm{h.c.} \quad\text{on leg B},
\end{equation}

Applying the conjugation $U^\dagger(\cdot)U$ again to the boundary terms, we realize the following for both legs:
\begin{itemize}
\item Leg $A$: 
Under conjugation, $b_{1,A} \to b_{1,A}$, since the exponent has an empty sum ($s<1$), so it is unchanged, while $b_{L,A} \to b_{L,A}e^{+i\theta\sum_{s<L}n_{s,B}}$. Therefore, the leg A boundary term (Eq.~\eqref{Eq:boundary_termA})transforms as
\begin{align}
&b_{1,A}^\dagger\,e^{+i\theta n_{L,B}}\,e^{+i\theta\sum_{s<L}n_{s,B}}\,b_{L,A} + \mathrm{h.c.} \nonumber \\
& \qquad =b_{1,A}^\dagger\,e^{+i\theta N_B}\,b_{L,A} + \mathrm{h.c.}
\end{align}

\item Leg $B$: 
$b_{1,B}\to e^{i\theta\sum_{r>1}n_{r,A}}b_{1,B}$, and $b_{L,B}$ has an empty sum ($r>L$), so it is unchanged. Therefore, the leg B boundary term (Eq.~\eqref{Eq:boundary_termB}) transforms as
\begin{align}
&b_{1,B}^\dagger\,e^{-i\theta\sum_{r>1}n_{r,A}}\,e^{-i\theta n_{1,A}}\,b_{L,B} + \mathrm{h.c.} \nonumber \\
&\qquad =b_{1,B}^\dagger\,e^{-i\theta N_A}\,b_{L,B} + \mathrm{h.c.}
\end{align}

\end{itemize}

Therefore, in the bulk the strings ``telescope'' to a single local density, while on the wraparound bonds they instead add up to the \emph{total} opposite-leg number, which does not cancel.

The Hamiltonian therefore maps to two decoupled hard-core rings $H^{\mathrm{hc}}(\varphi)$, with their statistical fluxes fixed by the total occupation number of an opposite ring. Let this hardcore Hamiltonian for a single leg be
\begin{equation}
H_{\ell}^{\mathrm{hc}}(\varphi) =-t\Bigl[\sum_{j=1}^{L-1}b_{j+1,X}^\dagger b_{j,X}+e^{i\varphi}b_{1,X}^\dagger b_{L,X}+\text{h.c.}\Bigr],
\end{equation}
which comprises the bulk term for OBC and the wraparound term for PBC for both legs $ \ell \in\{\mathrm{A,B} \}$, where $\varphi_A=\theta N_B$ and $\varphi_B=-\theta N_A$. The conjugated ladder Hamiltonian therefore becomes 
\begin{equation}
H(\theta) \equiv H^{\mathrm{PBC}}_{\mathrm{dec}}(\theta) = H_A^{\mathrm{hc}}(0; \theta N_B) + H_B^{\mathrm{hc}}(0; -\theta N_A),
\end{equation}
which is Eq.~\eqref{eq:hc_fluxes}. The eigenvalues of $H(\theta)$ are therefore sums $E_A+E_B$ of single-ring energies, and its eigenvectors are $U_{\mathrm{stat}}(\theta)\,|\psi_A\rangle\otimes|\psi_B\rangle$, with $|\psi_\ell\rangle$ an eigenvector of $H^{\mathrm{hc}}_\ell(0; \varphi_\ell)$.

Each leg is a hard-core chain, which the Jordan--Wigner map \cite{JordanWigner1928,LiebSchultzMattis1961} turns into free fermions, with the twist $\Phi_\ell$ of Eq.~\eqref{eq:fermion_twists} on a ring. The eigenvectors $|\psi_\ell\rangle$ are therefore Slater determinants of single-particle waves, and the energy is the sum of the occupied $-2t\cos k$. Under OBC the waves are $\sin[\pi mj/(L+1)]$, and Eq.~\eqref{eq:sector_equiv} gives the ladder eigenvectors $U_{\mathrm{stat}}(\theta)|\psi_A\rangle\otimes|\psi_B\rangle$ with $\theta$-independent energies. The ground state fills the lowest $N_\ell$ waves and is nondegenerate. Under PBC the waves are $e^{ikj}$ with $k=(2\pi n+\Phi_\ell)/L$, and the ground state of leg $\ell$ fills the $N_\ell$ consecutive momenta nearest $k=0$. A flux only shifts this block, relative to zero flux, by a common $\alpha_\ell$. Since $e^{i(k+\alpha)x}=e^{i\alpha x}e^{ikx}$, the amplitude of particles at $x_1<\dots<x_N$ is multiplied by $\prod_be^{i\alpha_\ell x_b}$, which is a product of on-site phases,
\begin{equation}
\label{eq:flux_phase}
|\psi_\ell(\varphi_\ell)\rangle\propto e^{i\alpha_\ell\sum_jj\,n_{j,\ell}}|\psi_\ell(0)\rangle,\qquad\alpha_\ell=-\frac{\delta_\ell}{L},
\end{equation}
where $\varphi_\ell=2\pi m_\ell+\delta_\ell$ with $m_\ell\in\mathbb Z$ and $\delta_\ell\in(-\pi,\pi]$. This holds away from the level crossings at $\delta_\ell=\pi$, the first of which is at $|\theta|=\pi/\max(N_A,N_B)$ for $0<N_A,N_B<L$. The phases $V=\prod_\ell e^{i\alpha_\ell\sum_jj\,n_{j,\ell}}$ act on each leg separately and commute with $U_{\mathrm{stat}}$, so the ground state is $|\Psi_0(\theta)\rangle\propto V\,U_{\mathrm{stat}}(\theta)|\Phi_0\rangle$, with $|\Phi_0\rangle=|\psi_A(0)\rangle\otimes|\psi_B(0)\rangle$ the zero-flux product state. Its occupation probabilities and leg-cut Schmidt values are therefore those of $U_{\mathrm{stat}}(\theta)|\Phi_0\rangle$. For both boundary conditions, $\theta$ enters the entanglement spectrum only through the statistical dressing. Under PBC it also enters the energy, through the fluxes.

\section{Spatial rung-cut spectrum under statistical dressing}
\label{App: trivial rung-cut}

For OBC, we prove that the statistical dressing cannot change the spectrum across a contiguous spatial cut in a fixed $(N_A,N_B)$ sector. Let $L_x$ denote the subsystem containing both legs on rungs $1,\dots,x$, and $R_x$ the subsystem containing both legs on rungs $x+1,\dots,L$. Then the string generator decomposes as
\begin{equation}
G=G_{L_x}+G_{R_x}+N_{A,R_x}N_{B,L_x},
\end{equation}
where
\begin{equation}
G_{L_x}\equiv \sum_{x\ge r>s} n_{r,A}n_{s,B},
\qquad
G_{R_x}\equiv \sum_{r>s>x} n_{r,A}n_{s,B},
\end{equation}
and $N_{A,R_x}=\sum_{r>x}n_{r,A}$ and $N_{B,L_x}=\sum_{s\le x}n_{s,B}$. Because $N_A=N_{A,L_x}+N_{A,R_x}$ is a fixed c-number in the sector, the only apparently nonlocal term can be rewritten as
\begin{equation}
N_{A,R_x}N_{B,L_x}=N_A N_{B,L_x}-N_{A,L_x}N_{B,L_x},
\end{equation}
which acts only on $L_x$. Since all number operators commute, this gives the exact factorization
\begin{equation}
U_{\mathrm{stat}}(\theta)=U^{\mathrm{sp}}_{L_x}(\theta)\otimes U^{\mathrm{sp}}_{R_x}(\theta),
\end{equation}
with
\begin{equation}
\begin{aligned}
U^{\mathrm{sp}}_{L_x}(\theta)&\equiv
\exp\!\left[i\theta\bigl(G_{L_x}+N_A N_{B,L_x}
-N_{A,L_x}N_{B,L_x}\bigr)\right],\\
U^{\mathrm{sp}}_{R_x}(\theta)&\equiv e^{i\theta G_{R_x}}.
\end{aligned}
\end{equation}
For OBC, the spatial-cut reduced state is therefore related to its $\theta=0$ counterpart by a local unitary conjugation on the retained half,
\begin{equation}
\rho_{L_x}(\theta)=U^{\mathrm{sp}}_{L_x}(\theta)\,\rho_{L_x}(0)\,U^{\mathrm{sp}}_{L_x}(\theta)^{\dagger},
\end{equation}
and its Schmidt values and entanglement spectrum are exactly independent of $\theta$. Under PBC the same factorization holds, and by Eq.~\eqref{eq:flux_phase} the induced fluxes change a nondegenerate ground state only by on-site phases. Its spatial-cut Schmidt values are therefore also $\theta$ independent.

\section{Numerical validation method}
\label{app:numerical_validation}

For each validation case, fixed-number occupation bases are enumerated exactly. The physical Hamiltonian is constructed from Eq.~\eqref{eq:Hladder}, including the wraparound bond for PBC. The decoupled Hamiltonian uses zero flux for OBC and the hard-core phases in Eq.~\eqref{eq:hc_fluxes} for PBC. A product eigenstate of the two decoupled chains is dressed directly by $U_{\mathrm{stat}}$, thereby avoiding arbitrary eigenvectors in a degenerate physical eigenspace.

The reduced state is computed independently by direct partial trace, the random-unitary sum in Eq.~\eqref{eq:rho_exact_fcs}, and the determinant in Eq.~\eqref{eq:det_form}. The first-two-cumulant channel is evaluated from exact one- and two-point occupation cumulants. The trace distance is
\begin{equation}
D_{\mathrm{tr}}(\rho,\sigma)=\frac12\|\rho-\sigma\|_1.
\end{equation}
The cases in Table~\ref{tab:validation} use nondegenerate mapped product eigenstates. Minimum eigenvalues are computed from the Hermitian part of each reduced state.

For the purity test, both covariance matrices in Eq.~\eqref{eq:purity_coefficient} are evaluated in the undressed product state. The finite-difference estimate uses $\theta=10^{-3}$, for which the relative discrepancy is governed by the $O(\theta^4)$ correction to the purity, or $O(\theta^2)$ after division by $\theta^2$.

\end{appendix}

\bibliography{refs}

\end{document}